\documentclass[twocolumn,showpacs,superscriptaddress,preprintnumbers,amsmath,amssymb,epsfig,floatfix,prx]{revtex4-1}
\usepackage{amsmath}
\usepackage{multirow}
\usepackage{amsfonts}
\usepackage{amssymb}
\usepackage{calligra}
\usepackage{calrsfs}
\DeclareMathAlphabet{\mathcalligra}{T1}{calligra}{l}{m}
\usepackage{epsfig}
\usepackage{graphicx}
\usepackage{dcolumn}
\usepackage{color}
\usepackage{natbib}  
\usepackage{hyperref}
\usepackage{breakurl}
\hypersetup{colorlinks=true, citecolor=blue, urlcolor=blue, linkcolor=blue}
\usepackage{bm}
\usepackage{booktabs}
\newcolumntype{C}{>{$}c<{$}}
\AtBeginDocument{
\heavyrulewidth=.08em
\lightrulewidth=.05em
\cmidrulewidth=.03em
\belowrulesep=.65ex
\belowbottomsep=0pt
\aboverulesep=.4ex
\abovetopsep=0pt
\cmidrulesep=\doublerulesep
\cmidrulekern=.5em
\defaultaddspace=.5em
}

\newcolumntype{L}[1]{>{\raggedright\arraybackslash}p{#1}}
\newcolumntype{C}[1]{>{\centering\arraybackslash}p{#1}}
\newcolumntype{R}[1]{>{\raggedleft\arraybackslash}p{#1}}

\begin{document}
\title{van der Waals heterostructure for photocatalysis: Graphitic carbon nitride and Janus transition-metal dichalcogenides}
\author{Srilatha Arra}
\affiliation{Department of Chemistry, Indian Institute of Science Education and Research, Pune 411008, India}
\author{Rohit Babar}
\affiliation{Department of Physics, Indian Institute of Science Education and Research, Pune 411008, India}
\author{Mukul Kabir}
\email{Corresponding author: mukul.kabir@iiserpune.ac.in} 
\affiliation{Department of Physics, Indian Institute of Science Education and Research, Pune 411008, India}
\affiliation{Centre for Energy Science, Indian Institute of Science Education and Research, Pune 411008, India}
\date{\today}

\begin{abstract}
Converting solar energy into chemical energy by splitting water is a promising means to generate a sustainable and renewable solution without detrimental environmental impact. The two-dimensional semiconductors serve as potential catalysts in this regard, and here we combine Janus transition-metal dichalcogenides (MoXY, X/Y = S, Se, Te) and graphitic carbon nitride in a van der Waals heterostructure. Within the first-principles calculations, we investigate the electronic, optical and excitonic properties that determine the photocatalytic activity. Due to the internal electric field, the photogenerated electrons and holes are separated in the MoXY layers, and also generates high overpotentials for the redox reactions. The high optical absorptions span throughout the entire visible and near ultraviolet regime in these heterostructure nanocomposites. Further, the lower exciton binding, calculated within the two-dimensional hydrogenic model, indicates efficient charge separation. Enormous tunability of photocatalytic properties in such heterostructures should attract considerable theoretical and experimental attention in future.     
\end{abstract}
\maketitle

\section{Introduction}
The fascinating mechanical, electronic and optical properties of two-dimensional (2D) materials and their concurrent tunability have opened up its applications in the modern electronic,  catalytic, and in energy conversion and storage devices.~\cite{Bonaccorso1246501, ncomms8873, C6NR00546B, acs.chemrev.6b00558, adma.201704548} Combining properties in 2D heterostructures provides further tunability in this context. Among other energy applications, the photocatalytic water splitting on the 2D materials has been attracted particular attention  to improve quantum efficiency. Apart from the easily tuneable intrinsic electronic structure, the 2D materials pose several advantages over other forms of nanomaterials. It maximizes the surface area for optical absorption and photocatalytic reaction. Further, it minimizes the distance travelled by the photogenerated electron and holes reducing recombination. In contrast, high exciton binding in the two-dimension due to reduced electron screening is a significant concern.~\citep{PhysRevLett.113.076802, nmat4061, nnano.2015.71, PhysRevB.99.045432} Despite significant achievements in optimizing photocatalytic activity, the low quantum efficiency hinders possible practical applications and warrants further material optimization.     

Several 2D materials such as metal oxides and hydroxides,~\citep{cm203293j, C2EE24148J, C4TA02678K, C5NR09248E, C6CS00343E} metal chalcogenides,~\citep{ncomms2066, ange.201204675, nmat3700, acsnano.5b04979} and metal-free semiconductors~\citep{nmat2317,science.aaa3145,adma.201303611,SHE2016138, C5EE03732H, acs.jpcc.7b12649} have been investigated as photocatalysts. The metal chalcogenides and metal-free graphitic carbon nitride (g-C$_3$N$_4$) have attracted particular attention. The free-standing SnS$_2$ and wide-gap ZnSe single-layers exhibit much enhanced photocurrent, water splitting efficiency and photostability.~\citep{ange.201204675, ncomms2066}  The transition-metal dichalcogenides (TMDCs) offer encouraging optical properties in the visible region along with exceptionally high carrier mobility, which makes them excellent candidate materials  for photocatalysis.~\citep{nmat3700, acsnano.5b04979} On the other hand, heptazine based g-C$_3$N$_4$ is the most promising metal-free candidate, which show photocatalytic activity under visible-light irradiation, however, it suffers from extremely low quantum efficiency.~\citep{nmat2317} In general, the catalytic activity in these materials is further tuned by doping,~\citep{adma.201303611, SHE2016138, anie.201209017,C5EE02650D} thickness,~\citep{ SHE2016138, jp061874w, ncomms2066} defect engineering~\citep{ adfm.201503221, aenm.201600436} and making composites including 2D van der Waals heterostructures.~\citep{anie.201210294,ange.201410172,ZHANG2015298,ncomms11480, acscatal.5b02036} For example, the quantum efficiency has been drastically improved for the g-C$_3$N$_4$ quantum dots and while modified with iodine, oxygen, phosphorous and boron.~\citep{science.aaa3145,adma.201303611,SHE2016138,anie.201209017,C5EE02650D} 
 
The 2D van der Waals stacking is an attractive strategy to combine materials with distinctive properties to optimize the photocatalytic properties. Apart from the improved light absorption,  the effective inter-layer charge separation can trigger reduction and oxidation of water on the different component layers.   The concept was developed by combining MoS$_2$ and g-C$_3$N$_4$ in a heterostructure, which showed improved catalytic activity.~\citep{ anie.201210294} Since then several MoS$_2$ and g-C$_3$N$_4$ based heterostructures have been proposed.~\citep{ange.201410172,ZHANG2015298,ncomms11480, acscatal.5b02036} 

\begin{figure*}[!t]
\begin{center}
\rotatebox{0}{\includegraphics[width=0.98\textwidth]{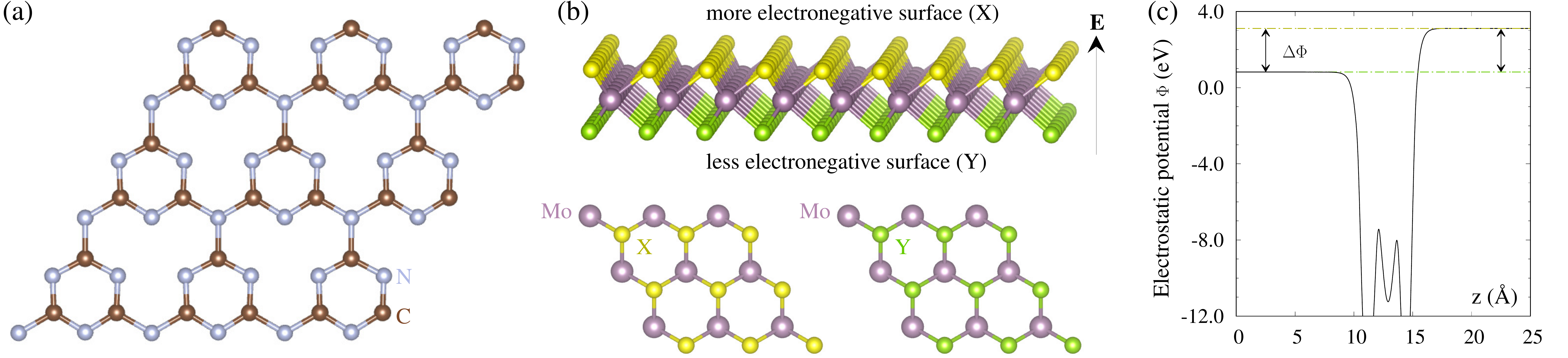}}
\caption{(a) The hexagonal 2D lattice of triazine-based graphitic carbon nitride (TGCN). No corrugation is observed in the free-standing single-layer TGCN.  (b) The side and top views for the single-layer Janus MoXY (X/Y = S, Se and Te), where the out-of-plane crystal symmetry is broken, and an internal electric field $\mathbf{E}$ is developed due to the difference in Pauling electronegativity. (c) Schematic description of a high electrostatic potential difference $\Delta \Phi$ that is thus produced between the X and Y surfaces.
}
\label{fig:figure1}
\end{center}
\end{figure*}

The breaking of out-of-plane structural symmetry in Janus TMDC provides another effective way to manipulate the electronic and optical properties. Single-layer MoSSe has been synthesized recently through a chemical vapour deposition based selenization or sulphurization methods.~\citep{nnano.2017.100, acsnano.7b03186} The presence of the optically active vertical dipole presents an interesting light-matter interaction in this symmetry-broken TMDC. On the other hand, depending on the size of the nitrogen-linked aromatic moieties there are two structural models for g-C$_3$N$_4$ $-$ triazine (C$_3$N$_4$) and heptazine C$_6$N$_7$ based graphitic sheets.  Thus, owing to their structural differences, the intrinsic electronic and optical properties vary substantially.~\citep{jpcc.5b07059} While enormous experimental attempts have been put forward for heptazine based g-C$_3$N$_4$,~\citep{nmat2317,science.aaa3145,adma.201303611,SHE2016138, C5EE03732H, acs.jpcc.7b12649}  it is only  recently that the triazine-based counterpart was successfully synthesized.~\citep{anie.201402191}

Here, we combine Mo-based Janus TMDCs and triazine-based g-C$_3$N$_4$ in a heterostructure architecture to investigate the properties that primarily determine the catalytic activity. Within the first-principles calculations the structural, electronic, and optical properties are studied comprehensively. Further, a two-dimensional hydrogenic exciton model is used to investigate exciton binding, where the parameters are derived from the first-principles calculations. Charge separation in the MoXY layers due to the internal electric field and lower exciton binding in these heterostructure nanocomposites indicate efficient charge separation favourable for photocatalysis.  Effective utilization of the solar spectrum and high overpotential drive the redox reactions efficiently.

\section{Methodology}
The first-principles calculations were carried out using the density functional theory as implemented in the Vienna {\em Ab Initio} Simulation Package.~\citep{PhysRevB.47.558,PhysRevB.54.11169} The wave function of the systems was described within the projector augmented wave formalism and expanded in plane-wave basis with 500 eV cut-off for the kinetic energy.~\citep{PhysRevB.50.17953} The exchange-correlation energy was treated with the Perdew-Burke-Ernzerhof functional form of generalized gradient approximation during the structural optimization.~\citep{PhysRevLett.77.3865} The weak van der Waals interaction was treated with non-local correlation functional vdW-DF-optB86.~\citep{PhysRevLett.92.246401,0953-8984-22-2-022201} All the structures were completely optimized until all the force components are less than 0.01 eV/\AA\ threshold. Subsequent electronic and optical properties are calculated using the hybrid Heyd-Scuseria-Ernzerhof (HSE06) exchange-correlational functional where a fraction of the exact Hartree-Fock exchange was considered.~\citep{10.1063/1.1564060} A $\Gamma$-centered 9$\times$9$\times$1, 17$\times$17$\times$1, and 4$\times$4$\times$1 $k$-mesh was used to sample the corresponding Brillouin zone for single-layers of g-C$_3$N$_4$, Janus MoXY (X/Y = S, Se, Te) and the van der Waals g-C$_3$N$_4$/MoXY heterostructures. We used a 30 \AA\ vacuum in the direction perpendicular to the surface to minimize the spurious periodic interaction between the images. Dipole correction was incorporated due to the presence of an intrinsic electric field in the Janus MoXY TMDCs.

\section{Results and Discussion}
First, we discuss the single layers of TGCN and Janus MoXY TMDCs. Next, we discuss the van der Waals MoXY/TGCN  heterostructures, and investigate the electronic and optical properties in the context of their photocatalytic abilities. Finally, we discuss exciton renormalization in the composite structures within a two-dimensional hydrogenic model.

\subsection{Single-layers of TGCN and Janus MoXY}
While the  heptazine based g-C$_3$N$_4$ has been studied extensively,~\citep{nmat2317,science.aaa3145,adma.201303611,SHE2016138, C5EE03732H} very little attention has been put forward for TGCN since it was successfully synthesized.~\citep{anie.201402191}  The calculated hexagonal in-plane lattice parameter of 4.78 \AA~ for the single-layer TGCN [Figure~\ref{fig:figure1}(a)] is consistent with the experimental bulk value and previous theoretical calculations.~\citep{anie.201402191,C4TA00275J} There are two distinct nitrogen sites in the lattice [Figure~\ref{fig:figure1}(a)]. The nitrogens that are bonded with two (three) neighbouring carbon atoms and form shorter 1.33 \AA\  (longer 1.46 \AA) C$-$N bonds. The HSE06 calculated bandstructure indicates the TGCN to be a direct gap semiconductor and the corresponding gap increases with the fractional Hartree-exchange $a_{\rm H}$. The calculated gap of 2.7 eV for $a_{\rm H}$=0.15 is consistent with the previous calculations,~\citep{anie.201402191,C4TA00275J} which increases to 3.3 eV for the conventional HSE06 functional with $a_{\rm H}$=0.25 (Table~\ref{table1}). In comparison, the experimental gap for the macroscopic TGCN flake measured via optical absorption was in the 1.6-2.0 eV range.~\citep{anie.201402191}  However, due to stronger quantum confinement in the single-layer, the bandgap is expected to be larger. The valence band is composed of the electrons from the undercoordinated nitrogens in the lattice, while the conduction band originates from the carbon and undercoordinated nitrogen atoms. In contrast, the electronic states corresponding to the three-coordinated nitrogens appear much deeper in energy.  

\begin{table}[!t]
\caption{Indirect and direct bandgaps $E_g^{i}$ and $E_g^{d}$ calculated using the HSE06 hybrid exchange-correlation functional for the single-layer MoSSe, MoSTe, TGCN, and their various heterostructures. The potential difference $\Delta \Phi$ across the two surfaces, and the calculated overpotentials $\chi_{_{{\rm H}_2}}$ and $\chi_{_{{\rm O}_2}}$  for the hydrogen and oxygen evolution reactions. All values are in eV.}
\begin{tabular}{L{2.5cm}C{1cm}C{1cm}C{1cm}C{1cm}C{1cm}} 
\hline
\hline
 System &  $E_g^{i}$ &  $E_g^{d}$ &  $\Delta \Phi$  & $\chi_{_{{\rm H}_2}}$   & $\chi_{_{{\rm O}_2}}$  \\ 
\hline
 MoSSe            &  $-$    &   2.06  &   0.78  &   0.56  &   1.08 \\
 MoSTe            &  1.63    &   1.88  &   1.64  &   1.06  &   0.95 \\
 TGCN             &  $-$   &   3.32  &   $-$   &   1.81  &   0.29 \\
 MoSSe/TGCN       &          &         &         &         &        \\
 S-facing         &  2.15    &   2.33  &   0.62  &   0.10  &   0.21 \\
 Se-facing        &  2.15    &   2.33  &   0.98  &   0.89  &   1.02 \\
 MoSTe/TGCN       &          &         &         &         &        \\
 S-facing         &  1.73    &   1.80  &   1.54  &   0.63  &   1.07 \\
 Te-facing        &  1.73    &   1.80  &   1.95  &   1.27  &   1.17 \\
\hline
\hline
\label{table1} 
\end{tabular}
\end{table}

The Janus MoSSe monolayer in Figure~\ref{fig:figure1}(b) was recently synthesized via selenization and sulphurization of MoS$_2$ and  MoSe$_2$, respectively. ~\citep{nnano.2017.100, acsnano.7b03186} The calculated lattice parameter of MoSSe, MoSTe and MoSeTe are found to be 3.23, 3.34, and 3.41 \AA,  which are consistent with the available experimental and previous theoretical results.~\citep{nnano.2017.100, acsnano.7b03186,acs.jpcc.7b11584,C7TA10015A,acs.nanolett.8b03474} 
The lattice parameters of MoXY is found to be the arithmetic mean of MoX$_2$ and MoY$_2$. Further, the Mo$-$X and  Mo$-$Y bonds in Janus MoXY are almost equal to the corresponding bonds in the pristine  MoX$_2$ and MoY$_2$. 

The Janus TMDCs retain the semiconducting properties with a bandgap equivalent to the average of their natural counterparts. The HSE06 gaps are calculated to be 2.06, 1.63 and 1.76 eV for MoSSe, MoSTe, and MoSeTe, respectively, and are consistent with the previous calculation.~\citep{acs.jpcc.7b11584}  
While MoSSe and MoSeTe are found to be direct-gap semiconductors with the gap at the K-point, the bandgap of MoSTe is found to be indirect in nature. While the conduction band minimum (CBM) in MoSTe is retained at the K-point,  the valence band maximum (VBM) moves to the $\Gamma$-point.  The out-of-plane structural symmetry in broken in the Janus MoXY and an internal electric field is developed due to the difference in electronegativity of the X and Y elements [Figure~\ref{fig:figure1}(b)]. Thus, a finite dipole moment perpendicular to the 2D surface is generated.~\citep{nnano.2017.100, acs.jpcc.7b11584} As a result, a large electrostatic potential difference $\Delta \Phi$ is developed between the X and Y surfaces [Figure~\ref{fig:figure1}(c)] and should be carefully incorporated while the VBM and CBM are normalized with the electrostatic potential far from the 2D surface. The calculated  $\Delta \Phi$  is 0.78, 1.64 and 0.89  eV for MoSSe, MoSTe, and MoSeTe, respectively.

\begin{figure}[!t]
\begin{center}
\rotatebox{0}{\includegraphics[width=0.47\textwidth]{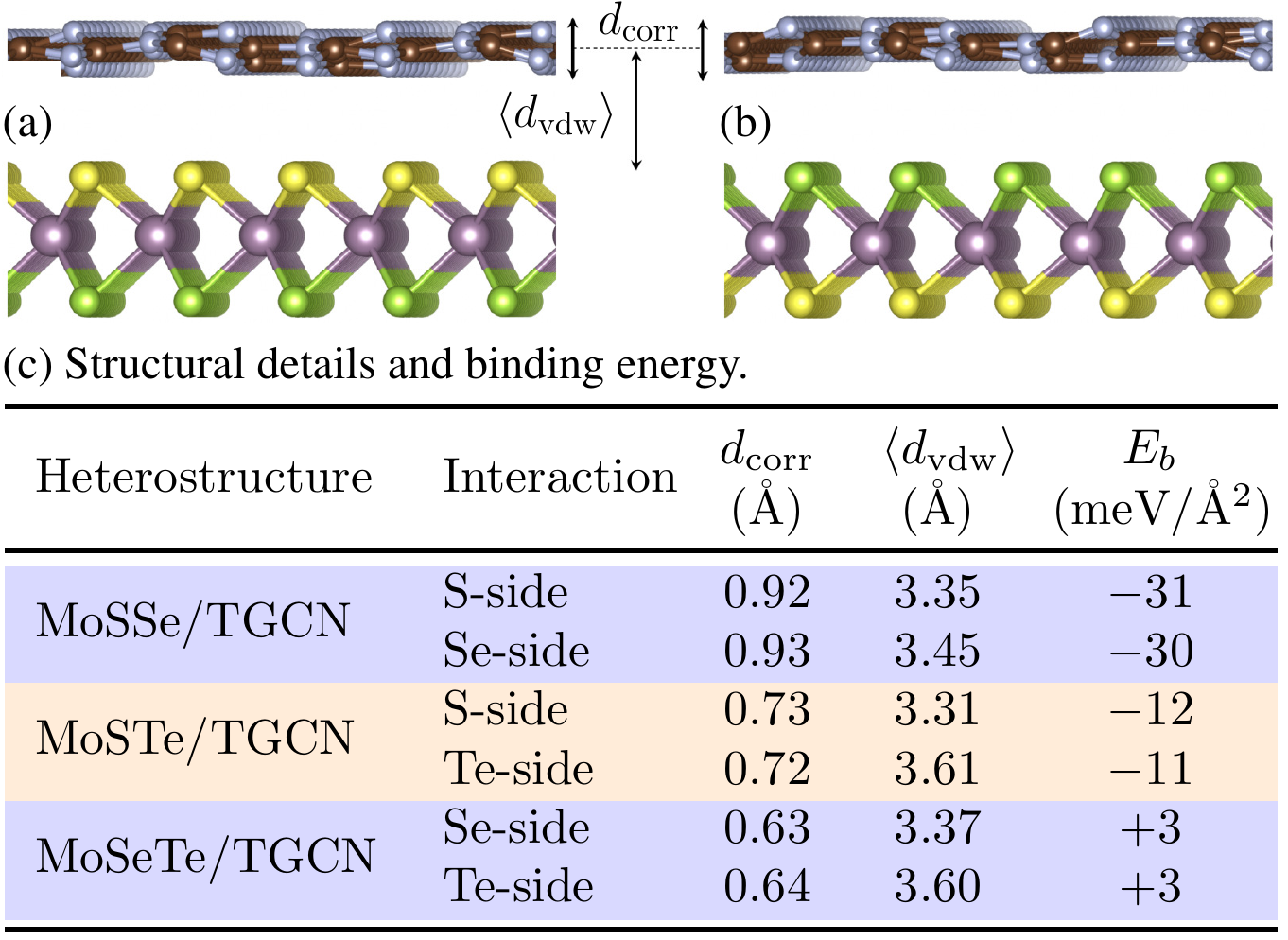}}
\caption{In the MoXY/TGCN heterostructures, the X or Y sublayer of Janus MoXY interacts with the TGCN. The MoSSe/TGCN heterostructure is shown for the (a) S-side  and (b) Se-side interactions. (c) The structural parameters, such as the corrugation in the TGCN layer $d_{\rm corr}$ and the average van der Waals separation $\langle d_{\rm vdw}\rangle$ between the layers, for different heterostructures and the corresponding binding energies. The $\langle d_{\rm vdw}\rangle$ increases with increasing chalcogen mass. The MoSeTe/TGCN heterostructures are not thermodynamically stable.}
\label{fig:figure2}
\end{center}
\end{figure}

\begin{figure*}[!t]
\begin{center}
\rotatebox{0}{\includegraphics[width=0.96\textwidth]{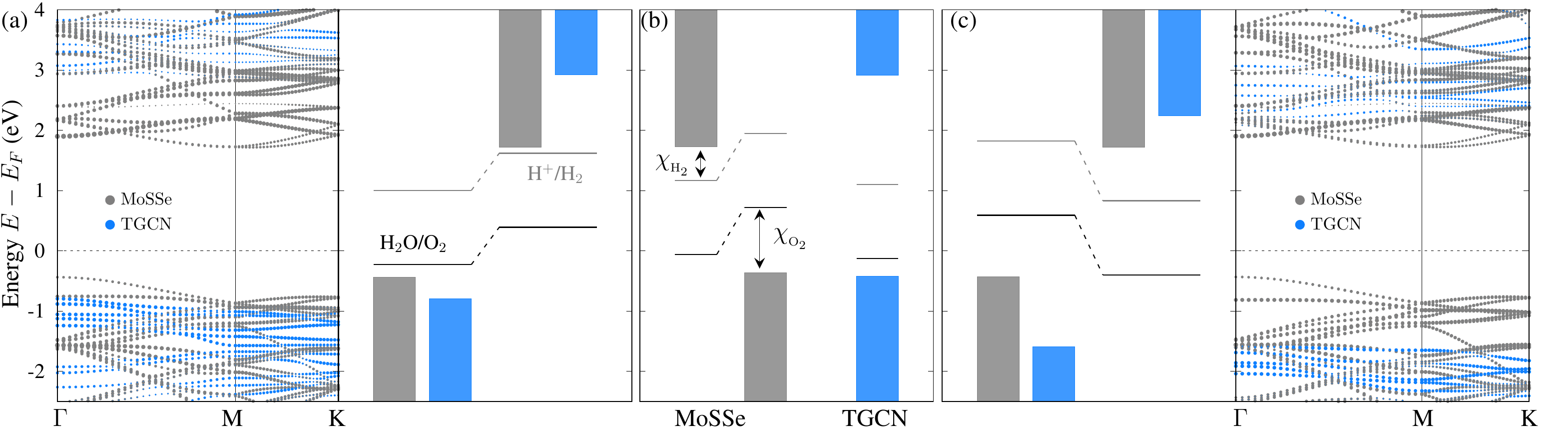}}
\rotatebox{0}{\includegraphics[width=0.96\textwidth]{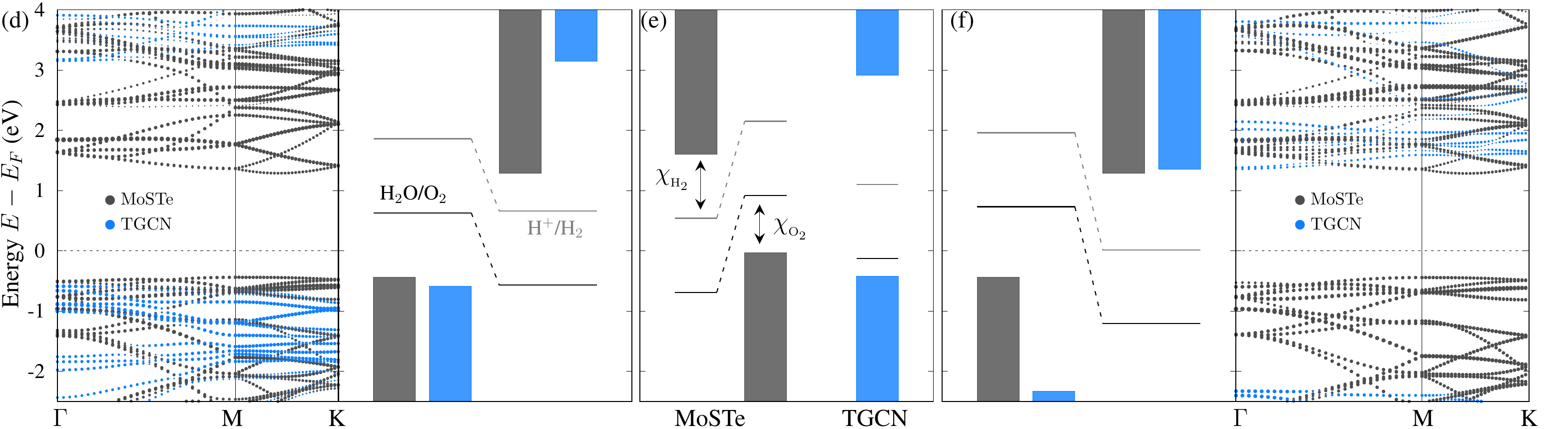}}
\caption{Band structure and band edge alignment for MoXY/TGCN heterostructures showing for  (a) S-side interaction and (c) Se-side interaction in MoSSe/TGCN, (d) S-side interaction and (f) Te-side interaction in MoSTe/TGCN. Band alignments for the single-layer MoSSe, TGCN, and MoSTe are shown in (b) and (e). The redox potentials of H$_2$O/O$_2$ and H$^+$/H$_2$ are represented with black and grey lines, respectively. The overpotential for oxygen and hydrogen evolution reactions $\chi_{_{{\rm O}_2}}$ and $\chi_{_{{\rm H}_2}}$ are indicated in (b) and (e). 
}
\label{fig:figure3}
\end{center}
\end{figure*}

\subsection{TGCN and Janus MoXY heterostructures}
While the single-layer TGCN structure is atomically flat [Figure~\ref{fig:figure1}(a)], a large corrugation $d_{\rm corr}$ is observed in the MoXY/TGCN heterostructures (Figure~\ref{fig:figure2}). We have considered both possible configurations, while the X or Y sublayer of MoXY interacts with the TGCN [Figure~\ref{fig:figure2}(a) and (b)]. The structure and the concurrent electronic properties are modified accordingly [Figure~\ref{fig:figure2}(c)]. The corrugation is higher for MoSSe/TGCN than in the MoSTe/TGCN and MoSeTe/TGCN heterostructures. The average van der Waals distance between the layers $\langle d_{\rm vdw} \rangle$ increases with increasing mass of the chalcogen [Figure~\ref{fig:figure2}(c)]. The heterostructures are thermodynamically stable except the MoSeTe/TGCN. The binding energy is calculated as $E_b = -[E({\rm MoXY/TGCN}) - E({\rm MoXY}) - E({\rm TGCN})]$ as shown in Figure~\ref{fig:figure2}(c).  We did not further investigate the MoSeTe/TGCN structures as they are not thermodynamically stable. 
The redox potentials for the oxygen evolution reaction (OER, H$_2$O/O$_2$) and hydrogen evolution reaction (HER, H$^+$/H$_2$) are determined by the relative electrostatic potential of the two respective surfaces that are normalized by the vacuum potential. The overpotentials indicate the redox abilities of photogenerated carriers and are defined as the potential difference between the VBM/CBM and the corresponding redox potentials. Therefore, the $\chi_{_{{\rm O}_2}}$ ($\chi_{_{{\rm H}_2}}$) is the energy difference between the VBM (CBM) and OER (HER) redox potential (Table~\ref{table1} and Figure~\ref{fig:figure3}).    

Before we discuss the electronic structure of these heterostructure nanocomposites  in the context of photocatalytic activity, we start with the photocatalytic activities of the single layers of TGCN and MoXY. Similar to the heptazine based counterpart, we find that the valence band maximum (VBM) and conduction band minimum (CBM) for the single-layer TCGN lie at energies such that the photo-generated electrons and holes catalyse water-splitting [Figure~\ref{fig:figure3}(b)]. However, an imbalance in the calculated overpotential is observed, $\chi_{_{{\rm O}_2}}= 0.29$ and $\chi_{_{{\rm H}_2}}=1.81$ eV (Table~\ref{table1}). The photocatalytic activity of the Janus MoXY have been discussed in literature,~\citep{acs.jpcc.7b11584,C7TA10015A} and have proposed advantages over the pristine MoX$_2$.  Due to the internal electric field shown in Figure~\ref{fig:figure1} (a) the fundamental restriction to the band gap ($>$ 1.23 eV) is lifted, and the redox potentials between the two surfaces shift by the electrostatic potential difference of $\Delta \Phi$ (Figure~\ref{fig:figure3}). (b) The photo-generated hot carriers are separated on the different X/Y sublayers indicating very efficient charge separation. For example, the photoexcited electrons and holes are accumulated on the Se and S side, respectively, for single-layer MoSSe. Therefore, the reduction and oxidation reactions are triggered on the different sides of the MoXY monolayer. The present results for these individual Janus MoXY are consistent with the previous results.~\citep{acs.jpcc.7b11584,C7TA10015A} Further, high overpotentials indicate efficient redox activity (Table~\ref{table1} and Figure~\ref{fig:figure3}). 

\begin{figure*}[!t]
\begin{center}
\rotatebox{0}{\includegraphics[width=0.96\textwidth]{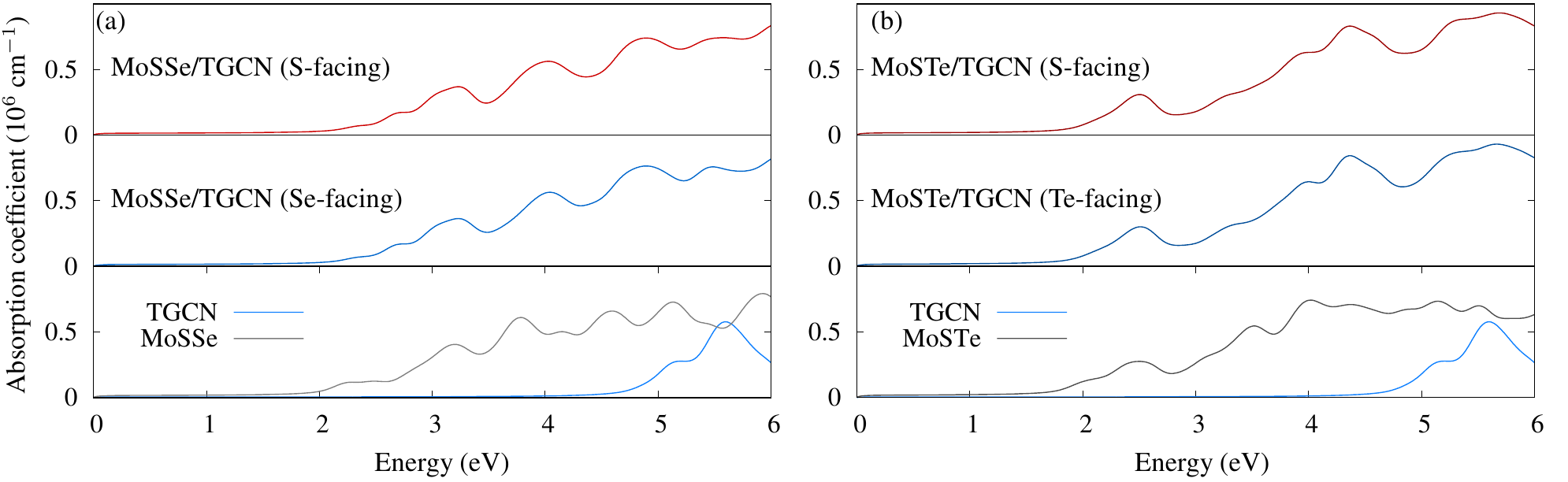}}
\caption{The optical absorption coefficient $\Lambda(\omega)$ calculated for the (a) MoSSe/TGCN and (b) MoSTe/TGCN heterostructure nanocomposites. All possible interaction architectures are considered. All the heterostructures show strong photoabsorption in the entire visible range and extend to the ultraviolet regime.}
\label{fig:figure4}
\end{center}
\end{figure*}

Having discussed the effectiveness of the single-layers of MoSSe, MoSTe and TGCN toward the redox reactions, we investigate their van der Waals heterostructures shown in Figure~\ref{fig:figure2}.  The band structure in MoSSe/TGCN heterostructures is altered in both the configurations in which S/Se sublayer interacts with the TGCN [Figure~\ref{fig:figure3}(a) and (c)]. In both cases, while the overall band structure becomes indirect with  a  gap of 2.15 eV, the direct gap at the $\Gamma$-point is slightly higher at 2.33 eV. The layer-projected band structures in Figure~\ref{fig:figure3}(a) and (c) indicate  type I band alignment, and  both the VBM and CBM lies in the MoSSe layer. While the MoSSe band structure in both the composite structures are similar to that of the isolated MoSSe monolayer, the same corresponding to the TGCN is substantially perturbed. Further, due to a different $\Delta \Phi$ in these two configurations, the redox potentials are shifted [Figure~\ref{fig:figure3}(a) and (c)], which results in different values for the $\chi_{_{{\rm O}_2}}$ and $\chi_{_{{\rm H}_2}}$ overpotentials (Table~\ref{table1}).  The overpotentials are much larger in the case of Se-side interaction than the S-side interaction, which indicates high redox ability (Table~\ref{table1} and Figure~\ref{fig:figure3}).  A deeper investigation into the electronic density of states reveals that both VBM and CBM is composed of the Mo-$d_{xy}$ and Mo-$d_{z^2}$ orbitals in S/Se-facing MoSSe/TGCN configurations.  The charge density analysis indicates that due to the internal electric field, the photoexcited electron and holes are separated on the Se and S sublayers, respectively. Thus, while for the MoSSe/TGCN heterostructure with S-side interaction [Figure~\ref{fig:figure3}(a)], the excited electrons on the Se-side are exposed to the  H$^+$/H$_2$ reaction, the H$_2$O/O$_2$ redox reaction is blocked as the photogenerated holes at the S-side cannot be accessed. Conversely, for the heterostructure with Se-side interaction [Figure~\ref{fig:figure3}(c)], the H$_2$O/O$_2$ reaction takes place at the exposed S-sublayer, while the H$^+$/H$_2$ reaction is blocked.         

In the cases of S-faced and Te-faced MoSTe/TGCN heterostructures, both the MoSTe and TGCN band structures are perturbed, and the band gap remains indirect (1.73 eV) similar to that of the single-layer MoSSe (Table~\ref{table1} and Figure~\ref{fig:figure3}). However, the direct gap at the M-point is found to be slightly larger at 1.80 eV (Table~\ref{table1}). We find the type I band alignment for MoSTe/TGCN configurations, while both VBM and CBM are composed with the Mo-$d_{z^2}$, Mo-$d_{xy}$ and Mo-$d_{x^2-y^2}$ orbitals. In the MoSTe/TGCN heterostructures, the photoexcited electrons (holes) are separated on the Te (S) sublayers of MoSTe, respectively. The overpotentials $\chi_{_{{\rm O}_2}}$ and $\chi_{_{{\rm H}_2}}$ for all the heterostructures, except for the S-faced MoSSe/TGCN is calculated to be higher than 0.6 eV, indicating their high redox abilities. The $\chi_{_{{\rm O}_2}}$ and $\chi_{_{{\rm H}_2}}$ are much higher than the recently predicted M$_2$X$_3$ (M = Al, Ga, In; X = S, Se, Te) family of photocatalysts.~\citep{acs.nanolett.8b02561}

\subsection{Optical properties}
The frequency-dependent complex dielectric tensor $\varepsilon (\omega)$ = $\varepsilon{'} (\omega)$ + i$\varepsilon{''} (\omega)$ is used to calculate the optical properties. The imaginary part $\varepsilon{''} (\omega)$  is evaluated in the long-wavelength $\mathbf{q} \rightarrow \mathbf{0}$ limit,~\citep{PhysRevB.73.045112}
\begin{eqnarray}
\varepsilon''_{\alpha \beta}(\omega) & = & \frac{4\pi^2e^2}{\Omega} \lim\limits_{q\rightarrow 0}\frac{1}{q^2}\sum\limits_{c,v,\mathbf{k}}2w_{\mathbf{k}} \delta(\epsilon_{c\mathbf{k}} - \epsilon_{v\mathbf{k}} - \omega) \nonumber \\
&\times& \langle u_{c\mathbf{k}+\mathbf{e}_{\alpha}q} | u_{v\mathbf{k}} \rangle \langle u_{c\mathbf{k}+\mathbf{e}_{\beta}q} | u_{v\mathbf{k}} \rangle^*. 
\end{eqnarray}
The factor 2 inside the summation accounts for the spin degeneracy, $\Omega$ is the volume of the primitive cell, and $\omega_{\mathbf k}$ are $k$-point weights. The $\epsilon_{c\mathbf{k}}$ ($\epsilon_{v\mathbf{k}}$) are ${\mathbf k}$-dependent conduction (valence) band energies, $u_{c\mathbf{k}, v\mathbf{k}}$ are cell periodic part of the pseudo-wave-function, and $\mathbf{e}_{\alpha, \beta}$ are unit vectors  along the Cartesian directions. 
The real part $\varepsilon{'} (\omega)$ of the frequency dependent complex dielectric tensor $\varepsilon (\omega)$ is calculated using the Kramers-Kronig transformation, and the absorption co-efficient is calculated as $\Lambda_{\alpha \alpha} (\omega) = \frac{2\omega}{c}[|\varepsilon_{\alpha \alpha} (\omega)| - \varepsilon{'}_{\alpha \alpha} (\omega)]^{\frac{1}{2}}$. 

We calculate the absorption coefficient $\Lambda(\omega)$ in Figure~\ref{fig:figure4} using the HSE06 exchange-correlation functional, which means that the electron-hole interaction is not considered. However, we will estimate the binding between the photogenerated $e-h$ pair in the next section.  Both MoSSe and MoSTe show strong photoabsorption in the entire visible range (Figure~\ref{fig:figure4}). While single-layer of MoSSe is a direct gap semiconductor, the direct and indirect gaps in the single-layer MoSTe are similar, reflecting in high photoabsorption.  Although TGCN is a direct gap semiconductor with 3.3 eV HSE06 gap, the absorption is negligible below 4.5 eV, which is in agreement with the experimental and previous theoretical calculations.~\citep{anie.201402191, C4TA00275J}  The negligible optical transition between the VBM and CBM at the $\Gamma$-point pushes light absorption to the ultraviolet range. The nanocomposites form type I heterojunction as discussed earlier, and the corresponding visible light absorption is similar to that of the MoSSe and MoSTe monolayers, with very high $\Lambda > 10^5$ cm$^{-1}$ (Figure~\ref{fig:figure4}). Further, compared to the TGCN monolayer, the heterostructures exhibit more efficient ultraviolet absorption.

\subsection{Exciton in heterostructures}
The Coulomb interaction between the photoexcited electron and hole pairs is attractive, and the strength of this interaction in the quasiparticle called exciton is an important quantity, which directly affects the photocatalytic efficiency. Lower exciton binding indicates an easy charge separation, and concurrently be consumed in the redox reactions. Due to the substantial reduction in electron screening in two-dimension, the exciton binding in 2D semiconductors is exceedingly enhanced. A theoretical investigation in this regard and within the first-principles approach is difficult since the formalism in which the many-body perturbation theory is coupled with the Bethe-Salpeter equation is computationally very expensive. This difficulty necessitates modelling exciton within  the Keldysh formalism,~\citep{Keldysh1979} where the parameters for exciton binding can be calculated using a much less expensive first-principles calculations.  Within the Keldysh formalism,  the hydrogenic effective exciton Hamiltonian is expressed as, $H_{\rm x} = -\frac{\hbar^2}{2\mu} \nabla_r^2 + V_{\rm 2D}(r)$, where $\mu$ is the exciton reduced mass, and $r$ is the electron-hole separation. The two-dimensional and non-locally screened electron-hole interaction is described using the Struve $H_0$ and Bessel $Y_0$ function as, $V_{\rm 2D}(r) = - \frac{e^2}{4(\varepsilon_1+\varepsilon_2)\varepsilon_0 r_0} [H_0(\frac{r}{r_0}) + Y_0(\frac{r}{r_0})]$, where $\varepsilon_1$  and $\varepsilon_2$ are  the  dielectric  constant  of  the  upper  and  lower  media;  and $\varepsilon_0$ is  the  vacuum  permittivity.~\citep{PhysRevB.99.045432,PhysRevB.91.245421,PhysRevB.95.235434} The screening length $r_0 = 2\pi\chi_{\rm 2D}$ with $\chi_{\rm 2D}$ the 2D  polarizability, which is calculated using the static dielectric constant $\varepsilon$ of the 2D system, $\varepsilon (L_{\rm v}) = 1 + 4\pi\chi_{\rm 2D}/L_{\rm v} $, where $L_{\rm v} $ is the transverse vacuum size.~\citep{PhysRevB.84.085406} The $\varepsilon$ is calculated from the real part of the complex dielectric tensor $\epsilon (\omega)$ at zero frequency. Thus, the present model following Keldysh formalism requires the effective masses of the photoexcited electron and hole along with the static dielectric constant. In the present calculations, these parameters are calculated using the hybrid HSE functional. Previously, we have used a similar approach to describe exciton binding in anisotropic phosphorene derivatives, which also correctly described the exciton renormalization in the few-layer phosphorene and heterostructures.~\citep{PhysRevB.99.045432}  

\begin{table}[!t]
\caption{The parameters for the two-dimensional hydrogenic exciton model, the effective carrier mass $m_e^*, m_h^*$, and the static dielectric constant $\varepsilon$ that is calculated using 30 \AA~ vacuum perpendicular the surface. Exciton binding  $E^b_{\rm x}$ in  the heterostructure nanocomposites show renormalization with an isotropic extension $\xi$ of 1 nm. }
\begin{center}
\begin{tabular}{L{2.1cm}C{1.1cm}C{1.1cm}C{0.8cm}C{1.2cm}C{1.0cm}} 
\toprule
\toprule
System &  $m_e^*/m_e$ &  $m_h^*/m_e$ &  $\varepsilon$  & $E^b_{\rm x}$ (eV)   & $\xi$ (\AA) \\ 
\midrule
 MoSSe/TGCN       &             &            &            &            &           \\
 S-facing                 &  0.63    &   0.68  &   3.81  &   0.51  &   9.20 \\
 Se-facing               &  0.50    &   0.68  &   3.81  &   0.50  &   9.52 \\
 MoSTe/TGCN        &             &            &            &            &           \\
 S-facing                 &  0.44    &   1.81  &   4.53  &   0.44  &   9.75 \\
 Te-facing                &  0.43    &   1.84  &   4.53  &   0.44  &   8.83 \\
\bottomrule
\bottomrule
\label{table2} 
\end{tabular}
\end{center}
\end{table}

To establish the applicability of the hydrogenic model of exciton for the composite systems, we first calculated the exciton binding for the single-layer MoXY by considering the explicit electron-hole interaction within the many-body perturbation-theory-based GW method plus Bethe-Salpeter equation (BSE) formalism,~\citep{PhysRev.139.A796,PhysRev.84.1232,PhysRevLett.80.4510,PhysRevLett.81.2312} and compare the results with the model calculations. Model calculations of 0.58 and 0.57 eV exciton binding for the Janus MoSSe and MoSTe are in excellent agreement with the present BSE calculations of 0.53 and 0.59 eV, respectively. Due to better electron screening in the MoXY/TGCN heterostructure nanocomposites, the calculated $E^b_{\rm x}$ shows strong renormalization indicating easier charge separation (Table~\ref{table2}), which is favourable for photocatalysis.  It is important to note that, while X/Y-facing interaction with TGCN differently impacts the electronic band structure leading to differential over-potentials $\chi_{_{{\rm H}_2}}$   and $\chi_{_{{\rm O}_2}}$ ( Table~\ref{table1} and Figure~\ref{fig:figure3}), however, it does not alter the $E^b_{\rm x}$. Further, owing to the type I band alignment, the exciton is localized in the MoXY layer with an extension $\xi$ about 1 nm.

\section{Conclusions}
Within the first-principles calculations, we have investigated the van der Waals Janus MoXY/g-C$_3$N$_4$ heterostructures in the context of photocatalysis. While the overall electronic band structure for the nanocomposites are type I, due to the internal electric field the photogenerated electrons and holes are separated in the two-sides of MoXY.  The high overpotentials indicate efficient hydrogen and oxygen evolution reactions. High optical absorption above 10$^5$ cm$^{-1}$ in the entire visible region including near ultraviolet regime indicates excellent utilization of the solar spectrum. Further, we have calculated the exciton binding within a simplistic two-dimensional hydrogenic model as the BSE approach is impossible for the nanocomposites studied here. Compared to the single-layer MoXY, in the heterostructures, we observe strong exciton renormalization due to increased electron screening indicating efficient charge separation. The present results open up an enormous tunability and design opportunity in van der Waals heterostructures.

\begin{acknowledgements}
We acknowledge the supercomputing facilities at the Centre for Development of Advanced Computing, Pune; Inter University Accelerator Centre, Delhi; and at the Center for Computational Materials Science, Institute of Materials Research, Tohoku University. M.K. acknowledges funding from the Department of Science and Technology through Nano Mission project SR/NM/TP-13/2016 and the Science and Engineering Research Board through EMR/2016/006458 grant.
\end{acknowledgements}


%

\end{document}